\documentclass{article}
\usepackage{graphics}

\begin{document}

\title{Matching Conditions in Atomistic-Continuum Modeling of
Materials}

\author{Weinan E$^1$ and Zhongyi Huang$^2$ \\
  $^1$Department of Mathematics and PACM, Princeton University and \\
  School of Mathematics, Peking University \\
  $^2$PACM, Princeton University and \\ Department of Mathematical
      Sciences, Tsinghua University}

\date{\today}

\maketitle

\begin{abstract}
A new class of matching condition between the atomistic
and continuum regions is presented for the multi-scale modeling
of crystals. They ensure the accurate passage of large scale information
between the atomistic and continuum regions and at the same time
minimize the reflection of phonons at the interface. These matching
conditions can be made adaptive if we choose appropriate weight 
functions. Applications to dislocation dynamics and friction between
two-dimensional atomically flat crystal surfaces are described.
\end{abstract}

\

Traditionally two apparently separate approaches have been used to
model a continuous medium. The first
is the continuum theory, in the form of partial differential
equations describing the
conservation laws and constitutive relations. This approach has been
impressively successful in a number of
areas such as solid and fluid mechanics. It is very efficient, simple
and often involves very few
material parameters. But it  becomes inaccurate for problems in which
the
detailed atomistic processes affect
the macroscopic behavior of the medium, or when the scale of the medium
is small enough that the continuum
approximation becomes questionable. Such  situations are often found
in studies of properties and defects of micro- or nano- systems and
devices. The
second approach is atomistic, aiming at  knowing the detailed behavior
of 
each individual atom using molecular dynamics
or quantum mechanics. This approach can in principle accurately model
the underlying physical
processes. But it is often times prohibitively expensive.

Recently an alternative approach has been explored that couples the
atomistic and continuum approaches.
\cite{Tadmor1,Tadmor2,TimK1,TimK2,Rudd1,Rudd2,Cai}. The main idea is to
use atomistic  modeling at
places where the displacement field varies on an atomic scale, and the
continuum approach elsewhere.
Two representative examples of such a coupled approach 
 are the quasi-continuum method \cite{Tadmor1,Tadmor2}
and the coarse-grained molecular dynamics \cite{Rudd1,Rudd2}. Both
methods have been successfully
applied to quasi-static examples, but extension to dynamic
problems has
not been straightforward. 
The main difficulty lies in the proper matching 
between the atomistic and continuum regions. Since the details of
lattice vibrations,
the phonons, which are an intrinsic part of the atomistic model, cannot
be represented at the continuum level,
conditions must be met that the phonons are not reflected at the 
atomistic-continuum interface. Since the atomistic region is expected
to be a small part of the computational domain, violation of this
condition
quickly leads to local heating of the atomistic region and destroys the 
simulation result. In addition, the matching 
between the atomistic-continuum interface has to be such that large
scale information is accurately transmitted in both directions.

The main purpose of this paper is to introduce a new class of matching
conditions between atomistic and
continuum regions. These matching conditions have the property that
they
allow accurate passage of large
scale (scales that are represented by the continuum model) information
between the atomistic and continuum regions and no
reflection of phonon energy to the atomistic region. These conditions
can also be used in pure molecular dynamics
simulations as the border conditions to ensure no reflection of phonons
at the boundary of the simulation.

For the sake of clarity we will first explain the main issues and ideas
on a simple problem: a
one-dimensional chain of particles coupled by springs:
\begin{equation}
\label{eq:1}
\ddot{u}_j=u_{j+1}-2u_j+u_{j-1}
\end{equation}
The spring constant is set to be 1.
After discretization in time, we have
\begin{equation}
\label{eq:2}
\frac{u^{n+1}_j-2u^n_j+u^{n-1}_j}{\Delta
t^2}=u^n_{j+1}-2u^n_j+u^n_{j-1}
\end{equation}
where $u^n_j$ is the displacement of the $j$-th particle at time
$t=n\Delta t$.

Equation (\ref{eq:2}) is supposed to be solved for all integers $j$.
Now
let us assume that we will
truncate the computational domain and only compute $u^n_j$ for $j\ge0$.
At $j=0$, we will impose a new
boundary condition to make sure that the phonons arriving from $j>0$
are
not reflected back at $j=0$.

The phonon spectrum for (\ref{eq:2}) is obtained by looking for
solutions of the type
$u^n_j=e^{i(n \omega \Delta t +j \xi )}$. This gives us the relation
\begin{equation}
\label{eq:3}
\frac1{\Delta t}\sin\frac{\omega\Delta t}2=\sin\xi
\end{equation}
At $j=0$, we replace (\ref{eq:2}) by
\begin{equation}
\label{eq:4}
u^n_{0}=\sum_{k,j\ge0}a_{k,j}u^{n-k}_j\; \qquad a_{0,0} = 0
\end{equation}
We would like to determine the coefficients $\{a_{k,j}\}$. For the
simple problem at hand, it is
possible to obtain analytical formulas of $\{a_{k,j}\}$ such that the
imposition of (\ref{eq:4})
together with the solution of (\ref{eq:2}) for $j>0$ reproduces
exactly the solution of
(\ref{eq:2}) if it was solved for all integer values of $j$, i.e. an exact
reflectionless
boundary condition can be
found. The details of this is given in \cite{EH}. This exact boundary
condition should be the same as
the one found numerically in \cite{Cai}. It represents the exact
Green's function for (\ref{eq:2}) which is nonlocal. However,
this procedure appears to be impractical for realistic models,
particularly
when the atomistic region moves with time
which is the case that interests us.
Therefore we will not  pursue this direction here.

A practical solution is to restrict (\ref{eq:4}) to a finite number of
terms and look for the
coefficients $\{a_{k,j}\}$ that minimize reflection. In order to do
this, let us look for solutions of
the type
\begin{equation}
\label{eq:5}
u^n_j=e^{i(n\omega \Delta t+j\xi )}+R(\xi)e^{i(n\omega \Delta t -j\xi
)}
\end{equation}
where $\omega$ is given by (\ref{eq:3}).
$R(\xi)$ is the reflection coefficient at wavenumber $\xi$. 
Inserting (\ref{eq:5}) into (\ref{eq:4}), we obtain
\begin{equation}
\label{eq:6}
R(\xi)=-\frac{\sum a_{k,j}e^{i(j\xi -k\omega\Delta t)}-1}{\sum
a_{k,j}e^{-i(j\xi +k\omega\Delta t)}-1}
\end{equation}
The optimal coefficients $\{a_{k,j}\}$ are obtained by
\begin{equation}
\label{eq:7}
\min\int^\pi_0 W(\xi) |R(\xi)|^2d\xi
\end{equation} 
subject to the constraint
\begin{equation}
\label{eq:8}
R(0)=0,R'(0)=0,\mbox{ etc.}
\end{equation}
Here $W(\xi)$ is a weight function, which is chosen to be $W(\xi) = 1$
in the examples below.

Condition (\ref{eq:8}) guarantees that large scale information is
transmitted accurately, whereas
(\ref{eq:7}) guarantees that the total amount of reflection is
minimized.
This procedure offers a lot of
flexibility. For example, instead of $\int^\pi_0|R(\xi)|^2d\xi$, we can
minimize the total reflection
over certain carefully selected interval. 
Another possibility is to choose the weight function to be
 the (empirically computed) energy spectrum. The
coefficients
$\{a_{k,j}\}$ may then change in
time to reflect the change of the
nature of the small scales. In
practice, we found it preferable to use
$\int^{\pi-\delta}_0|R(\xi)|^2d\xi$ with some small $\delta$,
instead of $\int^\pi_0|R(\xi)|^2d\xi$, in order to minimize the
influence of $\xi=\pi$ for which we
always have $R(\pi)=1$.

Let us look at a few examples. If in (\ref{eq:4}) we only keep the
terms 
involving $a_{1,0}$ and $a_{1,1}$, then imposing the condition $R(0)=0$
gives
\begin{equation}
\label{eq:9}
u^n_0 = (1 - \Delta t) u^{n-1}_0 +\Delta t u^{n-1}_1
\end{equation}
If instead we keep terms involving $a_{0,1}, a_{1,0}$ and $a_{1,1}$, we
can 
then impose both $R(0)=0$ and $R'(0) = 0$. This gives us
\begin{equation}
\label{eq:10}
u_0^n = u_1^{n-1} + \frac{1-\Delta t}{1+\Delta t}(u_0^{n-1} - u_1^n)
\end{equation}
Conditions of the type (\ref{eq:9}) and (\ref{eq:10}) are intimately
 related to the
absorbing boundary conditions proposed and analyzed in
\cite{Clayton,Eng}
for the computation of waves.
These conditions perform well for low wavenumbers but are less
satisfactory at high wavenumbers.

To improve the performance at high wavenumbers let us consider a case
that 
include terms with
$k\le2,j\le3$ and minimize
$\int^{\pi-\delta}_0|R(\xi)|^2d\xi$ (with $\delta=0.125\pi$) subject to
the
condition $R(0)=0$, the
optimal coefficients can be easily found numerically and are given by
\begin{equation}
\label{eq:11}
(a_{k,j})=\left(\begin{array}{ccc}
1.95264069 & -7.4207\times10^{-2} & -1.4903\times10^{-2} \\ [.15in]
-0.95405524 & 7.4904\times10^{-2} &
1.5621\times10^{-2}\end{array}\right)
\end{equation}
If instead we only include terms such that $k\le3,j\le2$, then
\begin{equation}
\label{eq:12}
(a_{k,j})=\left(\begin{array}{cc}
2.9524 & 1.5150\times10^{-2} \\ [.15in]
-2.9065 & -3.0741\times10^{-2} \\ [.15in]
0.9541 & 1.5624\times10^{-2}\end{array}\right)
\end{equation}
The resulting reflection coefficients $R$ are displayed in Figure 1.

\begin{figure} 
\begin{center}
\resizebox{3.5in}{!}{\includegraphics{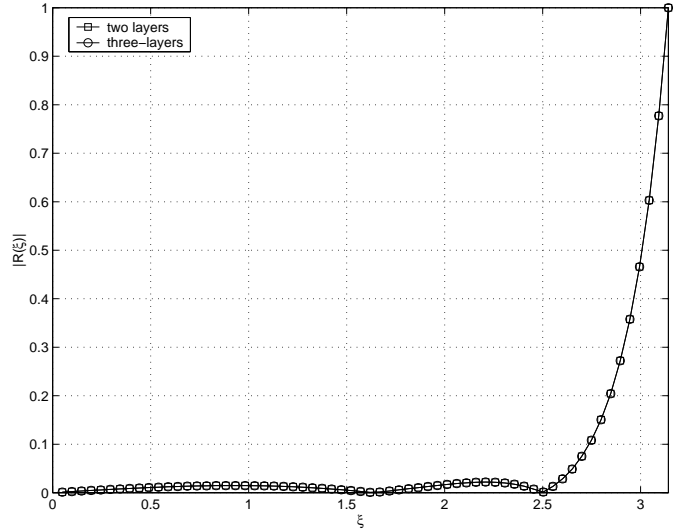}}
\caption{Reflection coefficients for (\ref{eq:11}) and (\ref{eq:12}).}
\end{center}
\end{figure}

We have applied this method to a number of problems. As the simplest
model that encompasses most of the
issues in a coupled atomistic/continuum simulation, we consider the
Frenkel-Kontorova model
\begin{equation}
\label{eq:13}
\ddot{x}_j=x_{j+1}-2x_j+x_{j-1}-U'(x_j)+f
\end{equation}
where $U$ is a periodic function with period 1, $f$ is an external
forcing. The continuum limit of this
equation is simply the Klein-Gordan equation
\begin{equation}
\label{eq:14}
u_{tt}=u_{xx}-Ku+f
\end{equation}
where $K=U''(0)$. We consider the case when there is a dislocation and
study its dynamics under a
constant applied forcing. We take $U(x)=(x-[x])^2$ where $[x]$ is the
integer part of $x$.
In this example we take (\ref{eq:13}) as our atomistic model, and 
(\ref{eq:14}) as our continuum model.
For the coupled
atomistic-continuum method, we use a standard second order finite
difference method for (\ref{eq:14})
in the region away from the dislocation, and we use (\ref{eq:13}) in
the
region around the
dislocation. However, we also place finite difference grid points in
the
atomistic region. At these points,
the values are obtained through averaging the values from the atomistic
model. At the
interface between the atomistic and
continuum regions, we decompose the displacement into a large scale and
a small scale part. The large scale part
is computed on the finite difference grid, using (\ref{eq:12}). The
small scale part is computed using
the reflectionless boundary conditions described earlier. The
interfacial
position between the MD and
continuum regions is moved adaptively 
according to an analysis of the wavelet
coefficients or  the local stress. 
Both strategies lead to similar results.
Care has to be exercised in order to restrict the size of the
atomistic region.
For example, when wavelet
coefficients are used in the criteria
to move the atomistic region, we found it more
efficient to use the intermediate levels of the wavelet coefficients
rather
than the finest level.

We first consider the case when a sharp transition is made between the
atomistic and continuum regions with a
1:16 ratio for the size of the grids. Figure 2 is a
comparison of the displacement
and velocity fields computed using the full atomistic  model and the
coupled
atomistic/continuum model, with $f=0.04$. The atomistic
region has 32 atoms. The full atomistic simulation has $10^4$. 
Dislocation appears as
a kink in the displacement field. Notice that at the
atomistic/continuum interface, there is still substantial phonon energy
which
is then suppressed by the
reflectionless boundary condition. 
No reflection of phonons back to the atomistic region is
observed.

We next consider a case with $f=0.02$, which alone is too weak to move
the dislocation, but to the left
of the dislocation, we add a sinusoidal wave to the initial data. The
dislocation moves as a
consequence of the combined effect of the force and the interaction
with
the wave. Yet in this case the
same atomistic/continuum method predicts an incorrect position for the
dislocation, as shown in Figure 3. The
discrepancy seems to grow linearly in time.
Improving the matching conditions does not seem to lead to significant
improvement.

\begin{figure}
\begin{center}
\resizebox{4in}{!}{\includegraphics{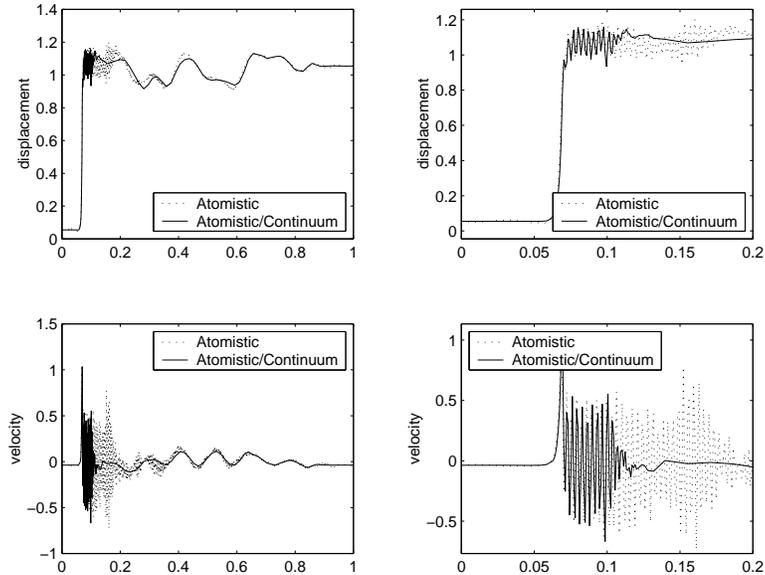}}
\caption{Comparison of the displacement and velocity profiles computed
using the full atomistic and the atomistic/continuum models, with $f=0.04$. 
The left column shows the results in the whole computational domain. 
The right column shows the details near the dislocation.}
\end{center}
\end{figure}

\begin{figure}
\begin{center}
\resizebox{4in}{!}{\includegraphics{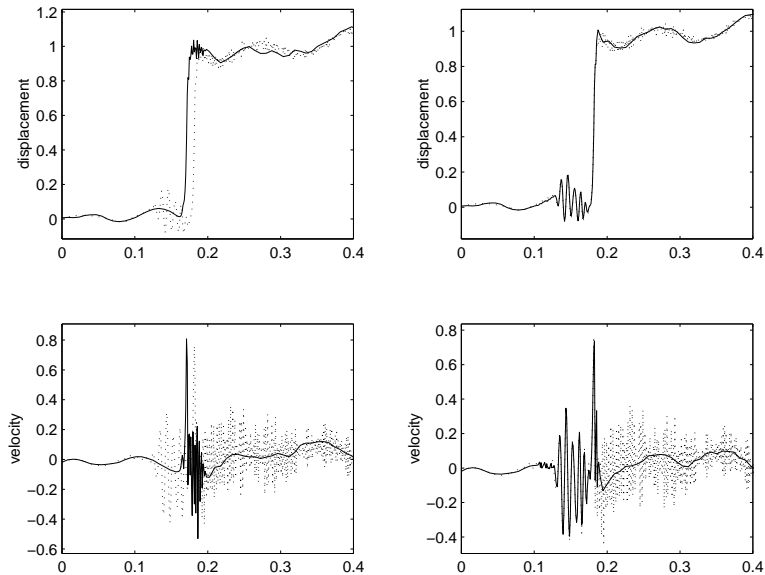}}
\caption{Comparison of the displacement and velocity profiles computed
using the full atomistic and the atomistic/continuum models, with $f=0.02$.
The left column shows the results when the transition
from the atomistic to continuum regions is sharp. The right columns
shows the results when the transition is gradual. Solid line is the
result of the atomistic/continuum method. The dash line is the result
of the full atomistic method. Only the region near the dislocation is shown.}
\end{center}
\end{figure}

The difference between this case and the case shown in Figure 2 is that
there is substantially more
energy at the intermediate scales.
This is clearly shown in the energy spectrum that we computed
for the two cases but it can also be seen in Figure 3 where
an appreciable amount of small
scale waves are present in front of the dislocation. Such
intermediate scales are suppressed in a
method that uses a sharp transition between the atomistic and continuum
regions. We therefore consider the
next alternative in which the atomistic/continuum transition is made
gradually
in a 1:2 or 1:4 ratio between
neighboring grids. The right column in Figure 3 shows the results of such a method that
uses
a gradual 1:2 transition. We
see that the correct dislocation position is now recovered.

Our second example is the friction between atomically flat crystal
surfaces.
To model this process atomistically, we use standard molecular dynamics
with
the Lennard-Jone potential \cite{Harrison,Robbins}. 
The two crystals are separated by a horizontal
interface. The atoms  in the bottom crystal are assumed to be much
heavier
than the atoms on top. A constant shear stress is applied at the top
surface.
The main issue here is how dissipation takes place. Physically the mean
kinetic energy is dissipated through conversion into phonons which then
convert into heat and exit the system. A standard practice in modeling
such
a process is to add a friction term to the molecular dynamics in order
to
control the temperature of the system \cite{Harrison,Robbins}. In contrast, we
ensure
the proper dissipation of phonons to the environment by imposing the
reflectionless
boundary conditions for the phonons. As a result we obtain a linear 
relationship between the mean displacement of the atoms in the top
crystal
as function of time, see Figure 4. The temperature of the system also 
saturates (Figure 4). Also plotted in Figure 4 is the result of the
mean displacement computed using the combined atomistic/continuum
method. Here the continuum model is the linear elastic wave
equation with Lame coefficients computed from the Lennard-Jones
potential.
The agreement between the full atomistic and the atomistic/continuum
simulation is quite satisfactory.

\begin{figure}
\begin{center}
\resizebox{4in}{!}{\includegraphics{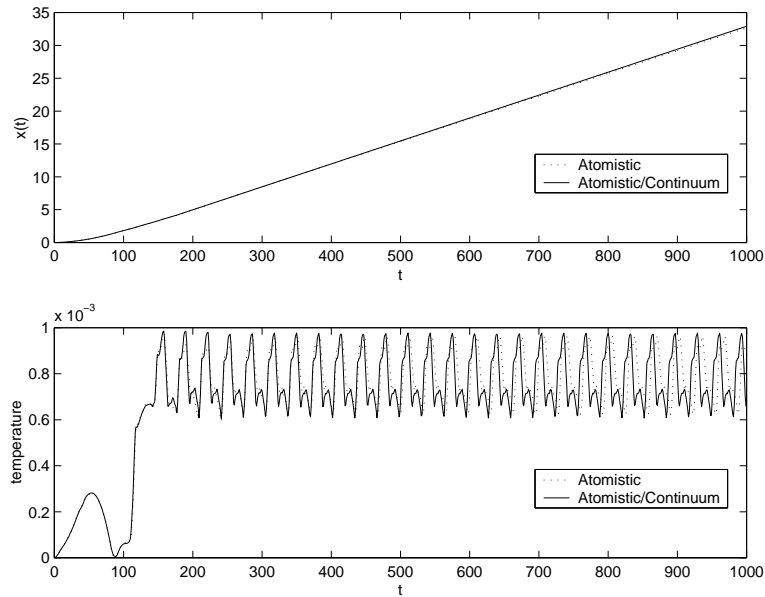}}
\caption{Displacement and temperature as a function of time for the friction
problem. Dash line is the result of the full molecular dynamics simulation.
The solid line is the result of the coupled atomistic/continuum method.}
\end{center}
\end{figure}

In conclusion, we presented a new strategy for the matching condition
at the atomistic/continuum interface in  multiscale modeling of
crystals.
These conditions are adaptive if we choose the weight functions in
(\ref{eq:7}) to reflect the evolving nature of the small scales. They
minimize the reflection of phonons and at the same time
ensure accurate passage of large scale information.

This work is supported by NSF through a PECASE award and by ONR grant
N00014-01-1-0674.

\end{document}